\documentclass[aps,mnras,twocolumn,showpacs,superscriptaddress,nofootinbib]{revtex4-1}  
\usepackage{graphicx}  
\usepackage{dcolumn}   
\usepackage{bm}   
\usepackage{float}
\usepackage[caption = false]{subfig}
\usepackage{amssymb}   
\usepackage{hyperref}
\usepackage{amsmath}
\usepackage{xcolor}
\usepackage{aas_macros}

\definecolor{green}{rgb}{0.2, 0.8, 0.2}
\definecolor{blue}{rgb}{0.0, 0.0, 1.0}

\newcommand{\ocen}{$\omega$-Cen\,}
\newcommand{\jfactor}{$J$-factor\,}
\newcommand{\gray}{$\gamma$-ray\,} 
\newcommand{\sigmav}{$\langle \sigma v \rangle$}

\begin{document}
\title{On the origin of the gamma-ray emission from Omega Centauri:\\ Milisecond pulsars and dark matter annihilation}

\author{Javier Reynoso-Cordova}
\email{reynosoj@fisica.ugto.mx}
\affiliation{
Departamento de F\'isica, DCI, Campus Le\'on, Universidad de
Guanajuato, 37150, Le\'on, Guanajuato, M\'exico}

\author{Oleg Burgue\~no}
\email{olegs@fisica.ugto.mx}

\affiliation{
Departamento de F\'isica, DCI, Campus Le\'on, Universidad de
Guanajuato, 37150, Le\'on, Guanajuato, M\'exico}

\author{Alex Geringer-Sameth}
\affiliation{Department of Physics, Imperial College London, Blackett Laboratory, Prince Consort Road, London SW7 2AZ, UK}

\author{Alma X. Gonz\'alez-Morales}
\affiliation{
Departamento de F\'isica, DCI, Campus Le\'on, Universidad de
Guanajuato, 37150, Le\'on, Guanajuato, M\'exico}
\affiliation{
Consejo Nacional de Ciencia y Tecnolog\'ia, Av. Insurgentes Sur 1582. Colonia Cr\'edito Constructor, Del. Benito Ju\'arez, C.P. 03940, M\'exico D.F. M\'exico}

\author{Stefano Profumo}
\affiliation{Department of Physics and Santa Cruz Institute for Particle Physics
University of California, Santa Cruz, CA 95064, USA}

\author{O. Valenzuela}
\affiliation{Instituto de Astronom\'ia, Universidad Nacional Aut\'onoma de M\'exico, A. P. 70-264, 04510, M\'exico, CDMX, M\'exico\\ }

\begin{abstract}
We explore two possible scenarios to explain the observed $\gamma$-ray emission associated with the atypical globular cluster $\omega$-Centauri: emission from millisecond pulsars (MSP) and dark matter (DM) annihilation. In the first case the total number of MSPs needed to produce the $\gamma$-ray flux is compatible with the known (but not confirmed) MSP candidates observed in X-rays. A DM interpretation is motivated by the possibility of $\omega$-Centauri being the remnant core of an ancient dwarf galaxy hosting a surviving DM component. At least two annihilation channels, light quarks and muons, can plausibly produce the observed $\gamma$-ray spectrum. We outline constraints on the parameter space of DM mass versus the product of the pair-annihilation cross section and integrated squared DM density (the so-called \jfactor). We translate upper limits on the dark matter content of $\omega$-Centauri into lower limits on the annihilation cross section. This shows s-wave annihilation into muons to be inconsistent with CMB observations, while a small window for annihilation into light quarks is allowed. Further analysis of $\omega$-Centauri's internal kinematics, and/or additional information on the resident MSP population will yield much stronger constraints and shed light about the origin of this otherwise mysterious $\gamma$-ray source.
\end{abstract}

\pacs{95.35.+d}

\maketitle

\section{Introduction}
\label{sec:Introduction}

Omega Centauri (\ocen) is a globular cluster (GC) that has received significant attention over the last few decades due to its unique properties. Arguably the most massive GC in the Galaxy with a mass $\sim 10^6 $ $\rm{M}_\odot$ \cite{2013MNRAS.436.2598W,Watkins:2015zza,McLaughlin:2006mp,2015ApJ...812..149W}, the nature of its stellar population is a rather hotly debated topic.  
Previous work suggests that \ocen's star formation history is quite different from that of other GCs, making it challenging to explain its stellar population as an isolated object \cite{Bekki:2005rx,Sollima:2005qy,Joo:2012gr,DAntona:2011ary,2017MNRAS.471L..31P}. \ocen has a high concentration of calcium and other heavy metals~\cite{Lee:2009gg} as compared to other GCs, which can be explained by supernovae metal enrichment \cite{Johnson:2009ku, 1995ApJS..101..181W}. However, \ocen's present gravitational potential is not deep enough to retain the ejected material from a supernova explosion \cite{Johnson:2009ku}. Such observations led to the hypothesis that this unique GC may be the remnant core of an ancient, tidally disrupted dwarf galaxy  \cite{1996ApJ...462..241N,Hilker:2000hr,Gnedin:2002un,Carraro:2000ji,Marcolini:2007kz,Romano:2007kh,2013ApJ...762...36J,Lee:2009gg,Bekki:2005rx,Rey:2003wk,Fellhauer:2004gc,Majewski:1999fw,1988IAUS..126..603Z,2004MNRAS.350.1141T}.  Unlike a typical GC, dwarf galaxies are dark matter (DM) dominated systems, as inferred from their dynamically-determined mass-to-light ratios \cite{Mateo:1998wg}. If \ocen is the vestige of one of such object it must have lost most of its external DM through tidal interactions with the Milky Way since its stellar kinematics has not yet shown any strong evidence for the presence of a DM halo. The recent discovery of tidal debris spread along \ocen's orbit adds evidence to this scenario~\cite{2019arXiv190209544I}.

The analysis of recent measurements of \ocen's internal kinematics \cite{2015ApJ...812..149W}, using the Hubble Space Telescope (HSP) proper-motion catalogue,  assuming a constant mass-to-light ratio $\Upsilon$, found $\Upsilon=2.66$. On the other hand, a value of $\Upsilon=1.87\pm 0.15$ was determined from stellar  population-synthesis modeling \cite{McLaughlin:2006mp}. Differences in mass-to-light ratios determined from the kinematics and stellar populations, could be explained due to variations in the initial mass function, or due to mass that can be attributed to the dark matter component, see the discussion in \cite{Cappellari:2011} for an example. Even though we do not claim here that the differences found for \ocen {\em per se} suggests the presence of a relic DM halo, it can certainly be used to derive an upper limit to the amount of DM present on it.

Dark matter is currently one of the most pressing puzzles in physics and astronomy, making up nearly 26\% \cite{Aghanim:2018eyx} of the total energy density in the Universe. With no non-gravitational signals detected so far, its fundamental nature remains a mystery. One distinct possibility is that particle dark matter annihilates with its anti-particle to produce stable Standard Model particles, including photons. The typical energies of such photons would then be around and below the dark matter particle mass. For weak-scale dark matter particles, this would entail the production of gamma rays, a possible tell-tale sign of dark matter annihilation.

In 2010 the Fermi Large Area Telescope (Fermi-LAT) collaboration~\cite{Abdo:2010nc,Abdo:2010dk,Ackermann:2012qk} published an analysis of $\gamma$-ray emission from a collection of Milky Way GCs. They reported a non-zero flux from some GCs including \ocen \cite{collaboration:2010bb}. The conventional explanation is that the emission is due to populations of millisecond pulsars (MSPs) within the GCs. These ancient, short-period pulsars are thought to be ``recycled'' neutron stars ``spun up'' through the loss of orbital angular momentum to their binary companion. For \ocen, the study concluded that a population of $19 \pm 9$ MSPs are needed to reproduce the data, assuming all MSPs have the same values for emissivity power and efficiency \cite{collaboration:2010bb}. Recent X-ray experiments show there are about thirty candidate MSPs in \ocen~\cite{Henleywillis:2018ufu}, but none of these has yet been confirmed as a pulsar. 

The search for high-energy particles, in particular $\gamma$-rays, produced by DM annihilation (or decay) in astrophysical systems has been extensively pursued over the last decade (for a review see \cite{profumo2017introduction,Conrad:2017pms}). Such an astronomical detection would shed light on the fundamental nature of DM and particle physics beyond the Standard Model. Given that \ocen could be the remnant core of a disrupted galaxy and actually have a (subdominant) DM component, it is timely to explore under which conditions DM annihilation can reproduce Fermi-LAT observations. This can impose important constraints on the parameter space of DM particle properties as well as on the distribution of DM within \ocen itself, e.g.~\cite{Bartels:2018qgr,Brown:2018pwq}. 

In this work we analyze $\sim$ 9.5 years of $\gamma$-ray observations by Fermi-LAT from the direction of \ocen. Data reduction and analysis methods are outlined in Sec.~\ref{sec::data}; In Sec.~\ref{sec:Flux} we present the details, and the assumptions we make, for the MSP and DM scenarios for the origin of the \gray emission; in Sec.~\ref{sec::analysis} we describe the results of the analysis in the two scenarios of MSP and DM and we also provide an extended discussion on the selection of the annihilation channels and possible constraints on the thermal averaged cross section.  Finally, we discuss our results and summarize our findings in Sec.~\ref{sec:Discussion}.

We emphasize that we do not claim here that DM annihilation is the most viable mechanism to explain the observed $\gamma$-ray emission from \ocen. Our goal is, rather, to place conservative constraints on  DM properties, including its abundance, in order to be compatible with observations. Finally we intend to compare this alternative explanation against a model where MSPs give rise to the $\gamma$-ray emission.

\section{Data selection and analysis}
\label{sec::data}
In this section we explain in detail the procedure used to reduce the data as well as the methodology to analyze it. The main purpose of the Fermi-LAT~\cite{Abdo:2010nc,Abdo:2010dk,Ackermann:2012qk} is to observe energetic $\gamma$-rays coming from Galactic and extra-galactic sources whose energy lies in the range from $\sim 100$ MeV up to $\gtrsim 300$ GeV.  As stated before, our object of study is the globular cluster Omega Centauri (\ocen), with coordinates: $201.6970$~RA and $-47.4795$~ DEC according to the Fourth Fermi Gamma-ray Source Catalog (4FGL) catalog \cite{Fermi-LAT:2019yla} ($2.89\times 10^{-5}$ deg from \ocen's position according to \cite{2000A&AS..143....9W}). Using v10r0p5 of the Fermi Science Tools \footnote{https://fermi.gsfc.nasa.gov/ssc/data/analysis/software/}, we select $\sim 9.5$ years of LAT data (between MET $239557417$s and $574181612$s). We adopt a region of interest (ROI) of $0.2 \deg$ radius around the position of \ocen in the 4FGL and select all Pass 8R3 Source events (\texttt{evclass=128} and \texttt{evtype=3}), in the $0.2-50$ GeV range within the ROI using \texttt{gtselect} with a maximum zenith angle of $90\deg$. We use the standard filter \texttt{"DATA\_QUAL$>0$ \&\& (LAT\_CONFIG$==1$)"} in \texttt{gtmktime}, we use \texttt{dcostheta=0.025} and \texttt{binsz=1} in \texttt{gtltcube}, and \texttt{theta=300} and \texttt{thetamax=1 deg} in \texttt{gtpsf} to obtain the exposure, $\epsilon$, (observation time times effective area) and point spread function, PSF, both functions of energy. We extract the measured energy spectrum in 15 equally log-spaced energy bins between 0.2 and 50~GeV using \texttt{gtbin} with the PHA1 algorithm.

 We analyze the data by fitting proposed models to the observed photon flux, see for instance~\cite{Abdo:2009jfa,Hooper:2016rap,Cholis:2014noa}  in the context of LAT data. 
 Since the the photon count coming from \ocen is low the observations are most accurately described by a Poisson likelihood, defined by 
\begin{equation}\label{eq:poisson}
\mathcal{L} = \prod_i \frac{e^{-\lambda_i}\lambda_i^{d_i}}{d_{i}!},
\end{equation}
where the product is over energy bins $i$, $d_i$ is the number of counts observed in each bin, and $\lambda_i$ is the corresponding model prediction. 

There is one known gamma-ray source (4FGL J1328.5-4727) within $1^\circ$ of our target according to the 4FGL\footnote{In the previous 3FGL catalog \cite{Acero:2015hja}, there were two nearby sources within 1 degree of \ocen. We report the analysis with only the one source present in the latest catalog. However, our conclusions are robust to the inclusion of the other one.} (besides the source associated with \ocen itself). According to the 4FGL, its energy spectrum corresponds to a power law with no cutoff. To take into account its contribution to the total number of events we will include it in the likelihood at the catalog position and use several nuisance parameters to describe its energy spectrum. 

The general form for the model is given by
\begin{equation}\label{eq::modelcomplete}
\lambda = M(E,\Theta) + S(E,\eta) + B(E),
\end{equation}
where $M$ corresponds to the emission from \ocen (to be defined in section \ref{sec::Flux_MSP} for MSP,  and \ref{sec::Flux_dm} for DM). $S$ corresponds to the nearby point source for which the flux is modeled with a simple power law with no cut-off: $d\phi/dE = N_{s}(E/E_p)^{-\alpha_s}$. Here $N_{s}$ and $\alpha_s$ are the normalization and the spectral index, respectively, and will be treated as nuisance parameters that we marginalize over ($E_p$ is an arbitrary reference energy which we set to 4726.70 MeV according to the 4FGL catalog). To obtain the number of events contributed by  this point source, we integrate the energy flux within the energy bins limits, as well as integrating the PSF (centered on the point source) over the \ocen ROI as specified in Eq.~\eqref{eq::int_PSF}.

Finally $B(E)$ corresponds to the diffuse and isotropic background components. We adopt the Fermi collaboration model\footnote{\url{http://fermi.gsfc.nasa.gov/ssc/data/access/lat/BackgroundModels.html}} consisting of interstellar galactic emission\footnote{\texttt{gll\_iem\_v07.fits}} and an isotropic component\footnote{\texttt{iso\_P8R3\_SOURCE\_V2\_v1.txt}}. The prediction $B(E)$ within each energy bin is found by multiplying the sum of these two component spectra by the LAT exposure and the solid angle of the ROI and then integrating over the energy bin.

With the likelihood defined, we perform a Monte Carlo Markov Chain (MCMC) analysis to generate posterior distributions for the model parameters. We use flat priors for parameters describing the emission from \ocen, whereas Gaussian priors are used for the those describing the nearby source. These are specified in section \ref{sec::Analysis_Results}.
For the MCMC we use the \texttt{emcee} package, which is an affine invariant ensemble sampler~\cite{2013PASP..125..306F}. Our base setup for the MCMC is to use 200 walkers with 1500 steps and a burn-in phase of 500 steps. To check for convergence we performed two checks: (a) we doubled the number of steps with a fixed number of walkers; (b) we doubled the number of walkers with the number of steps fixed as recommended by the authors of the sampler to reduce the noise in the chains. In both tests we find similar results to the base setup. Finally, to compare the different models we use the Akaike Information Criterion, $AIC= -2k-2\ln(\mathcal{Z})$, where $k$ is the number of free parameters in the model and $\mathcal{Z}$ is the evidence. Smaller values of the AIC indicate the model is a better fit to the data, after penalizing for the number of free parameters in the model. In this case evidences were computed using the \texttt{pymultinest} code \cite{Feroz:2007kg,2014A&A...564A.125B}, that uses a nested sampling algorithm, for which one of the main features is to compute the evidence. The use of \texttt{multinest} also served as a cross check of the \texttt{emcee} results.


\section{Possible explanations for the {\large $\omega$}-cen observed $\gamma$-rays events}
\label{sec:Flux}

\subsection{Milisecond Pulsars}
\label{sec::Flux_MSP}

Thus far, the only hypothesis explored to explain the observed flux of $\gamma$-rays from \ocen is that they originate from MSPs \cite{collaboration:2010bb,Cholis:2014noa,Hooper:2016rap}. In this scenario the  photon flux is commonly modelled with an exponentially cut-off power law given by 
\begin{equation}
\label{eq::exp_cutoff}
\frac{d\phi}{dE} = N_0 (E/{\rm MeV})^{-\Gamma} \mathrm{e}^{-E/E_{\rm cut}},
\end{equation}
where $N_0$ is the normalization, $\Gamma$ is the spectral index and $E_{\rm cut}$ is the energy cutoff. Those three are the free parameters to be fitted in our the analysis, together with two nuisance parameters associated to the nearby point source. To compute the predicted number of events detected in an energy bin with limits $E_1$ and $E_2$ we need to include information from the instrument response, namely the exposure $\epsilon$ and the point spread function (PSF) which gives the probability that a photon originating from \ocen is reconstructed within our defined ROI. Since these two quantities are energy-dependent, the expected number of events for the $i$-th energy bin is given by 
\begin{equation}
\label{eq::no_events_pulsar}
M(E_i)=\int_{E_i}^{E_{i+1}} \int_{\Delta \Omega}d \Omega  \hspace{1mm} \epsilon(E) \rm{PSF}(E,\Theta) \frac{d\phi}{dE}.
\end{equation}
Note that the above equation takes \ocen to be a point source of $\gamma$-rays. This is a valid assumption as the scale radius of the emission ought to be of the order the half-light radius \cite{Marcolini:2007kz}, which is less than the width of the PSF ($\gtrsim 0.2^\circ$) at the energies relevant to our analysis. 

Once the model parameters are determined, by means of the MCMC using the likelihood defined in Sec.~\ref{sec::data}, the next step is to compute the associated number of MSPs giving rise to the emission. Following \cite{collaboration:2010bb}, the number of MSPs is approximated by 
\begin{equation}
\label{eq::no_MSP}
N_{\rm{MSP}}= \frac{L_\gamma}{\left\langle \dot{E}\right \rangle \eta},
\end{equation}
where $L_\gamma$ is the isotropic luminosity, $\eta$ is the estimated average spin-down to $\gamma$-ray luminosity conversion efficiency and $\langle \dot{E} \rangle$ is the average spin-down power. We adopt the values of $\langle \dot{E} \rangle$ and $\eta$ from Ref.~\cite{collaboration:2010bb}. The simplest approach is to approximate $L_\gamma$ directly from the observed energy flux $S$ and the distance to the source $d$~\cite{collaboration:2010bb},
\begin{equation}
L_\gamma=4 \pi S d^2.
\label{eq:msp_luminosity}
\end{equation}
Note that Ref.~\cite{Hooper:2016rap} estimates the number of MSPs in a globular cluster by correlating the number of binary systems that can be formed given a stellar density and  $\gamma$-ray luminosity, whereas in Ref.~\cite{Cholis:2014noa} a correlation between the X-ray and $\gamma$-ray flux is used in order to compute how many MSPs could contribute to the $\gamma$-ray flux. These studies argue that the naive way to compute the number of MSPs used by \cite{collaboration:2010bb} can lead to a wrong answer since, for a globular cluster, the $\gamma$-ray emission could be due to one really luminous MSP or the sum of several fainter ones.  The details on how the MSPs are contributing to the $\gamma$-ray photon flux could lead to important differences in the analysis, specially when multiple components of the $\gamma$-ray source are considered. See for instance the analysis by \cite{Brown:2018pwq} and \cite{Bartels:2018qgr} on the 47 Tuc GC, which we discuss in section \ref{sec:Discussion}. As mentioned earlier there are $\sim 30$ sources with X-ray-MSP-like emission in \ocen \cite{Henleywillis:2018ufu}, but there is no confirmation with radio observations that these sources are pulsars. Therefore, we consider the simple approach given by Eq.~\eqref{eq:msp_luminosity} to be sufficient to obtain the order of magnitude of the number of MSPs required to explain \ocen's observed $\gamma$-ray photon flux.

\subsection{Gamma rays from annihilating dark matter}
\label{sec::Flux_dm}
As stated in Sec.~\ref{sec:Introduction}, one of our goals is to explore the possibility that \ocen's energy spectrum is produced by the annihilation of DM into Standard Model particles. In this case the  photon flux from \ocen is given by\footnote{Notice that Eq.~(\ref{eq::flux}) assumes that the dark matter is its own antiparticle; if this is not the case, and there is an equal number of dark matter particles and antiparticles, an additional factor 1/2 would be needed.} 
\begin{equation}
\label{eq::flux}
\frac{d\phi}{dE}= \frac{1}{8 \pi} J_{\Delta \Omega} \frac{\left\langle \sigma v \right\rangle}{m_\chi^2} \frac{dN}{dE},
\end{equation}
where $ \left \langle \sigma v \right\rangle $ is the thermal averaged annihilation cross section, $m_\chi$ is the dark matter particle mass, $dN/dE$ is the photon spectrum produced for each annihilation, and $J_{\Delta \Omega}$ is the so-called astrophysical \jfactor. The \jfactor encodes information about the DM density distribution within \ocen. It is the integral over a solid angle $\Delta \Omega$ of the $J$-profile $dJ/d\Omega$, which determines the spatial morphology of $\gamma$-ray emission from DM annihilation. The $J$-profile is defined as the integral of the square of the dark matter mass density $\rho(r)$ along a line of sight that makes an angle $\theta$ with the direction toward the center of \ocen (e.g.~\cite{Geringer-Sameth:2014yza}):
\begin{align}
\label{eq::Jfactor}
\frac{dJ}{d\Omega}(\theta) &= \int \rho^2(r(\theta,\ell)) d\ell, \\
J_{\Delta \Omega} &= \int_{\Delta \Omega} \frac{dJ}{d\Omega} d\Omega.
\end{align}
One common approach is to use the observed stellar or gas kinematics to constrain the DM density distribution and then use this density profile to determine the \jfactor for a given target. For nearby dwarf spheroidal galaxies this is usually done by fitting the free parameters of a proposed DM density profile using the Jeans equation (e.g.~\cite{Strigari:2006rd,Bonnivard:2015xpq}). For Galactic Center studies, the proposed Milky Way DM density profile is normalized so that it matches the inferred DM density in the solar neighborhood, and is then extrapolated to the very inner regions~\cite{Abazajian:2012pn,TheFermi-LAT:2017vmf}. In dwarf spheroidals, propagating the uncertainties in the observed kinematics and the influence of different DM and velocity anisotropy profiles have yielded relatively constrained estimates for the \jfactor~\cite[e.g.][]{2015MNRAS.451.2524M,Geringer-Sameth:2014yza,Bonnivard:2015xpq}.

We are not aware of any previous work pursuing this kind of analysis for the case of \ocen, i.e. there are no specific constraints to the DM density profile that could be associated with it. Performing a Jeans (or similar) analysis of \ocen kinematics including a DM component is by itself an interesting possibility which, however, is out of the scope of this paper. 
For this work we use available information on the different estimates of \ocen's mass-to-light ratio $\Upsilon$ to set an upper limit on the dark matter mass possibly contained within \ocen's half-light radius, $r_h$, and use that to pinpoint a reasonable range for the \jfactor. We spell this out next. 
\begin{figure*}[!ht]
\centering
\subfloat{\includegraphics[width=18cm]{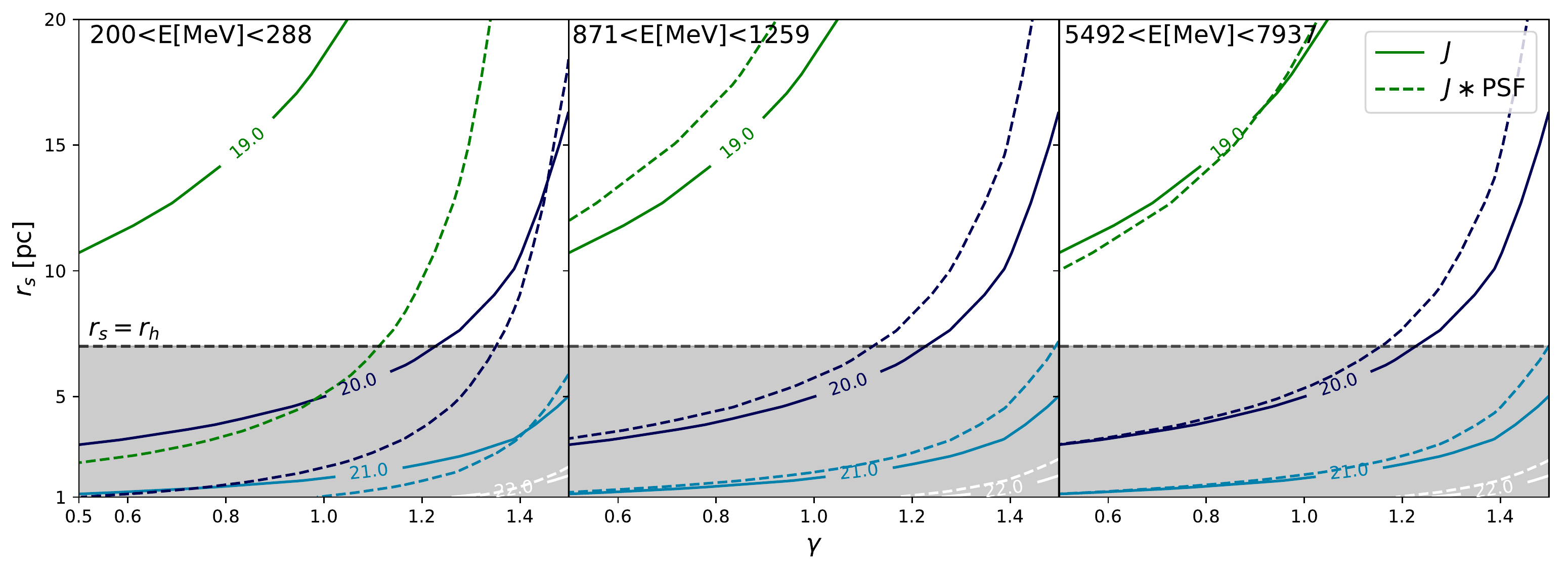}}
\caption{
Comparison between the \jfactor (Eq. \ref{eq::Jfactor}), in $\log_{10}$ scale, which is constant for all energy bins (solid lines)  and the convolution of the \jfactor with the PSF weighted by the PSF (Eq. \ref{eq::convolution}, dashed lines), as a function of the scale radius and inner slope $\gamma$ of the DM density profile (Eq.~\eqref{eq::NFW}). The three panels show results for three different energy bins. Shaded region below the $r_s=r_h$ would correspond to what we consider non-physical \jfactor values if  \ocen is the remnant of a disrupted galaxy.}

\label{fig:j_factor}
\end{figure*}
Using population-synthesis models to compute  stellar properties, a mass-to-light ratio of $\Upsilon_{\rm{syn}}=1.87\pm0.15$ was found in Ref.~\cite{McLaughlin:2006mp}. On the other hand, in Ref.~\cite{Watkins:2015zza}, spherical, non rotating, dynamical models with constant mass-to-light ratio were fitted to the most recent Hubble Space Telescope proper motions \cite{2014ApJ...797..115B}, finding $\Upsilon_{\rm{dyn}}=2.66\pm 0.04$. The tension between these two values might be due to differences in the initial mass functions (IMF)~\cite{Watkins:2015zza}. However the possibility that a dark matter component can explain the difference between dynamical and stellar mass-to-light estimates has not been discussed. We do not attempt to demonstrate here that the difference is minimized by including DM as another $\omega-$cen component, but merely use the difference between the two determinations of  $\Upsilon$ to set a possible maximum value for the DM contribution to the mass.

Following Ref.~\cite{Watkins:2013fna}, we compute the stellar mass using a cylindrical stellar density profile given by a Gaussian decomposition:

\begin{equation}
\label{eq::stellar_density}
\rho_*(z,R)=\sum_{j}^{M} \frac{M_j}{(2\pi \sigma_j^2)^{3/2} q_j} \rm{exp}\left[-\frac{1}{2 \sigma_j^2}\left( R^2 + \frac{z^2}{q_j^2} \right)\right],
\end{equation}
where $M_j$ are the different mass components, $\sigma_j$ controls the extent along the major axis, and $q_j$ is the projected flattening. We set all parameters as in~\cite{Watkins:2013fna}. Using this decomposition we compute the stellar mass present in \ocen for the two $\Upsilon$ values described above. Then we compute the difference between these two values which is approximately 
$\delta M(r_h)=1.01\times 10^5 \,{\rm M}_{\odot}$, where $r_h$ is the \ocen half light radius equal to 7 pc \cite{Watkins:2013fna}. We take $\delta M(r_h)$ as the maximum total mass within $r_h$ that could be attributed to a DM halo.  Note that according to the simulations of \cite{Bekki:2003qw}, the dark matter mass that could remain in a \ocen like object after a tidal disruption, assuming \ocen is the remnant of a dwarf galaxy, would be $M_{DM}\sim 4 \times 10^{5} M_\odot$. Since to date there are no observational constraints on \ocen's DM component we can only point out that the total mass limit we adopt is similar to what is expected from predictions based on these numerical simulations.

To model the DM density in Eq.~\eqref{eq::Jfactor}, we adopt a generalized Navarro-Frenk-White density profile;
\begin{equation}
\label{eq::NFW}
\rho (r)= \frac{\rho_s}{\left(\frac{r}{r_s} \right)^\gamma \left(1 + \left(\frac{r}{r_s}\right) \right)^{(3 - \gamma)}},
\end{equation}
where $\rho_s$ and  $r_s$ are the characteristic density and scale radius respectively.

Further, we make some assumptions on the properties of the DM halo. A simple picture of \ocen being the remnant of a disrupted galaxy would require the scale radius to be larger than the observed luminous half mass radius, otherwise the stellar component would not survive the disruption. On the other hand, one could extrapolate mass-concentration relations such as the one proposed in Ref.~\cite{Klypin2014:1411.4001v2}, which seems to hold even for small halos~\cite{Pilipenko2017:1703.06012v2}, and compute the scale radius for a halo mass of $10^6 M_{\odot}$, which results in a few parsecs. Therefore we can safely assume the scale radius to be larger or of the order of $r_{\rm h}$. For a given $r_s$ and $\gamma$ we determine $\rho_s$ so that the enclosed mass within the half-light radius is equal to the DM mass limit found from the difference between the values of the mass-to-light-ratios computed above: $M_{\rm DM}(r_h)=\delta M(r_h)=1.01\times 10^5\, {\rm M}_\odot$.

It is important to take into account that, just as in the calculation for the number of events using the MSP model, we must include the information regarding the instrument response. The case of DM is trickier than for MSPs since the angular extent of the dark matter emission profile may be comparable to or broader than the PSF. The way to incorporate the PSF information is through its convolution with the $J$-profile so that the observed number of events is given by
\begin{equation}
\label{eq::convolution}
\frac{dN}{dE}_{\rm observed}=\frac{1}{8 \pi}\frac{\left \langle \sigma v\right\rangle}{m_\chi^2}\frac{dN}{dE}\int_{\Delta \Omega}d\Omega \left(\frac{dJ}{d\Omega}*\rm{PSF}\right)(E,\theta).   \end{equation}

Over a small patch of sky the integral over solid angle becomes a 2D rectangular integral which can be performed using a zeroth order one-dimensional Hankel transform~\cite{Geringer-Sameth:2014yza}. For this we use the python package {\emph{mcfit}} which uses the FFTlog algorithm \cite{Hamilton:1999uv}. We give the details of this procedure in Appendix~\ref{sec::convolution}.

Formally, the density profile parameters  $r_h$ and $\gamma$ should be sampled by the MCMC, but this would involve performing the convolution on every MCMC step, which would be computationally expensive. As a simpler approach, we will use a single value of the \jfactor for all energy bins which is then multiplied by the energy-dependent PSF. Then we will give an estimate of the error that we might be introducing by this approximation. In Figure~\ref{fig:j_factor} we show, for three energy bins, how the convolution of the $J$-profile with the PSF integrated over the ROI (dashed lines), and divided by the integrated PSF, approaches to the unconvolved \jfactor for high energies (solid lines) and substantially differs for low energies (dashed lines). That is, for the high-energy bins the equality 
\begin{equation}
J_{\Delta \Omega} = \frac{\int_{\Delta \Omega} d\Omega \left( \frac{dJ}{d\Omega}\ast\rm{PSF}\right) (E,\theta)}{\int_{\Delta \Omega} d\Omega \rm{PSF(\theta)}}
\label{eq:comparison_j_factor}
\end{equation}
holds. The relatively large difference, between the convolved and unconvolved \jfactor in the low energy bins will lead to different results specifically for the analysis of the $\bar{q}q$ channel. As we discuss in Sec.~\ref{sec:Discussion:DM}, the  approximation made does not have a great impact on our conclusions. We also note that the analysis of \cite{deMenezes:2018ilq} prefers a point-source model of \ocen to an extended one, which is also consistent with our approximation of using a constant \jfactor for all energy bins.

\begin{figure}[!t]
\centering
\includegraphics[width=1\linewidth]{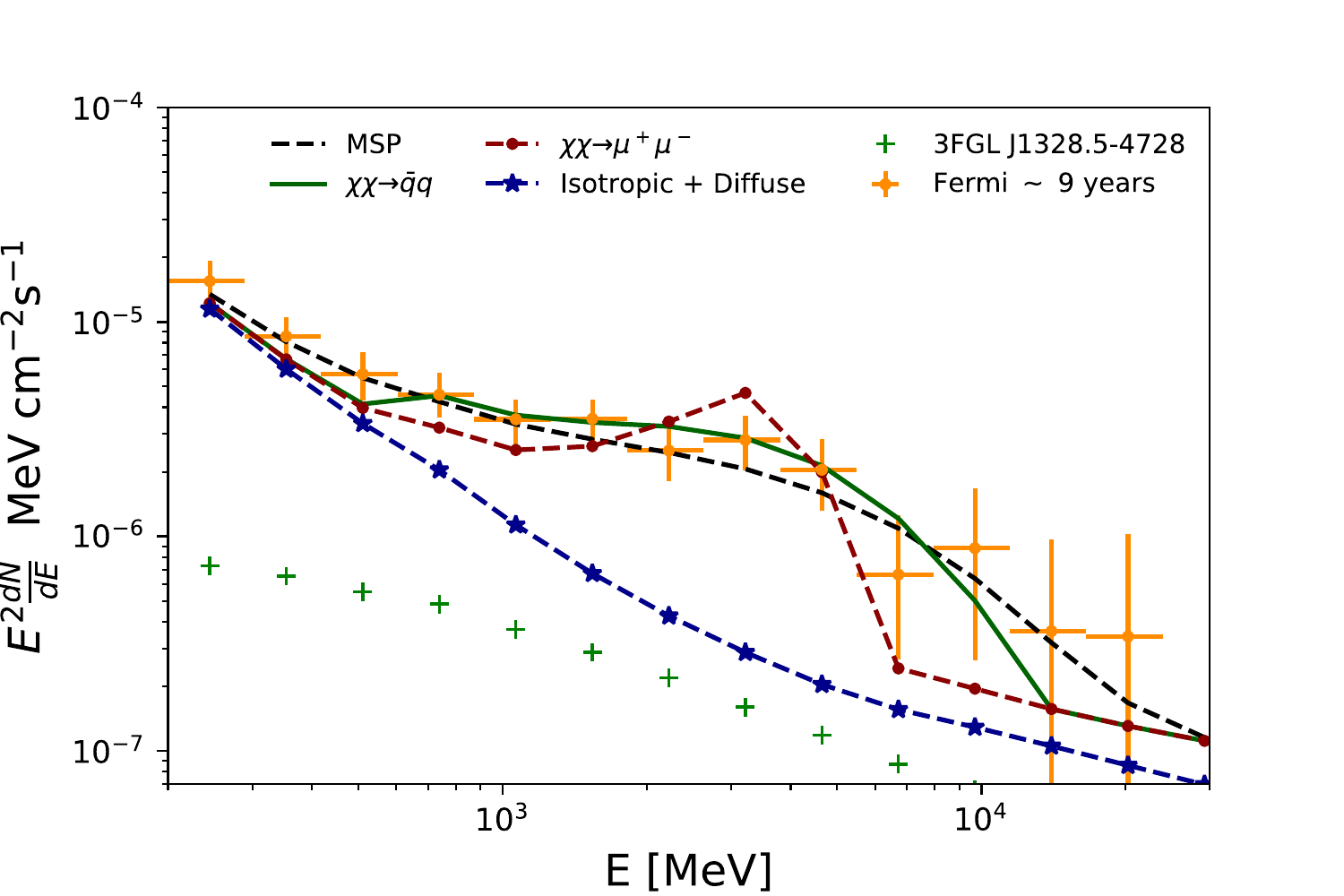}

\caption{ \ocen energy flux (yellow dots) as observed with the  Fermi-LAT corresponding to $\sim$ 9 years. Dashed black line shows the total flux predicted by the Millisecond Pulsar model (section \ref{sec::analysis_msp}). Green solid line and point-dashed line correspond to total flux for the DM models annihilating into light quark-antiquark $q\bar{q}$ and muon-antimoun $\mu^+ \mu^-$, respectively, as described on \ref{sec:analysis_DM}. For this plot we used the median values of the fitted parameters shown in Figure~\ref{fig::eventscomplete}. Also shown for reference is the background contribution to the total flux. }
\label{fig:flux}
\end{figure}

\section{Analysis and Results}
\label{sec::Analysis_Results}

Using the models for $\gamma$-ray emission from \ocen described in Sections~\ref{sec::Flux_MSP} and~\ref{sec::Flux_dm}, we now proceed to analyze the data using the statistical methods specified in Section~\ref{sec::data}.

\label{sec::analysis}
\subsection{MSPs}
\label{sec::analysis_msp}
\begin{table}
\renewcommand{\arraystretch}{1.5}
\begin{center}
\begin{tabular}{ c c c c } 
\boldmath{$\Gamma$} & \boldmath{$\log_{10}(E_{cut})$} & \boldmath{$\log_{10}(N_{0})$} & \boldmath{$N_{MSP}$} \\ 
 \hline
 $1.58^{+0.26}_{-0.26}$ & $3.55^{+0.18}_{-0.14}$ & $-6.89^{+0.71}_{-0.85}$ & $31.78^{+18.77}_{-7.97}$\\ 
 \hline
 \hline
  & \boldmath{$\log_{10}(N_{S})$} & \boldmath{$\alpha_{s}$} & AIC \\ 
 \hline
  & $-11.43^{+0.08}_{-0.10}$ & $2.70^{+0.13}_{-0.13}$ & 20.3 \\ 
\hline 
\end{tabular}
\caption{Median values obtained by the inference for the MSP model parameters (see figure~\ref{fig::eventscomplete}) and the corresponding Akaike information criterion parameter (AIC). The error bars for all the parameters are at a $95\%$ confidence.}
\end{center}
\label{table::MSP_parameters}
\end{table}

We model MSP emission in \ocen as a power law spectrum with exponential cut-off given by Eq.~\eqref{eq::exp_cutoff} with three free parameters: $N_0$, $\Gamma$, $E_{\rm cut}$. The two nuisance parameters $N_s$ and $\alpha_s$ describe the energy spectrum of the nearby source. For the \ocen parameters we use flat priors over the following ranges:
\begin{itemize}
    \item $0<\Gamma<3$,
    \item $2<\log_{10}(E_{\rm cut}/{\rm MeV})<8$,
    \item $-18<\log_{10}(N_0/{\rm MeV})<0$.
\end{itemize}
We adopt Gaussian priors for $N_s$ and $\alpha_s$ based on the best fitting values and error bars reported in the 4FGL catalog \cite{Fermi-LAT:2019yla}:
\begin{itemize}

\item $\dfrac{N_s}{10^{-14}} = 3.81 \pm 0.39 \;{\rm MeV}^{-1}$

\item $\alpha_s=1.98 \pm 0.07$
\end{itemize} 

\begin{figure*}[ht!]
\centering
\includegraphics[width=0.75\linewidth]{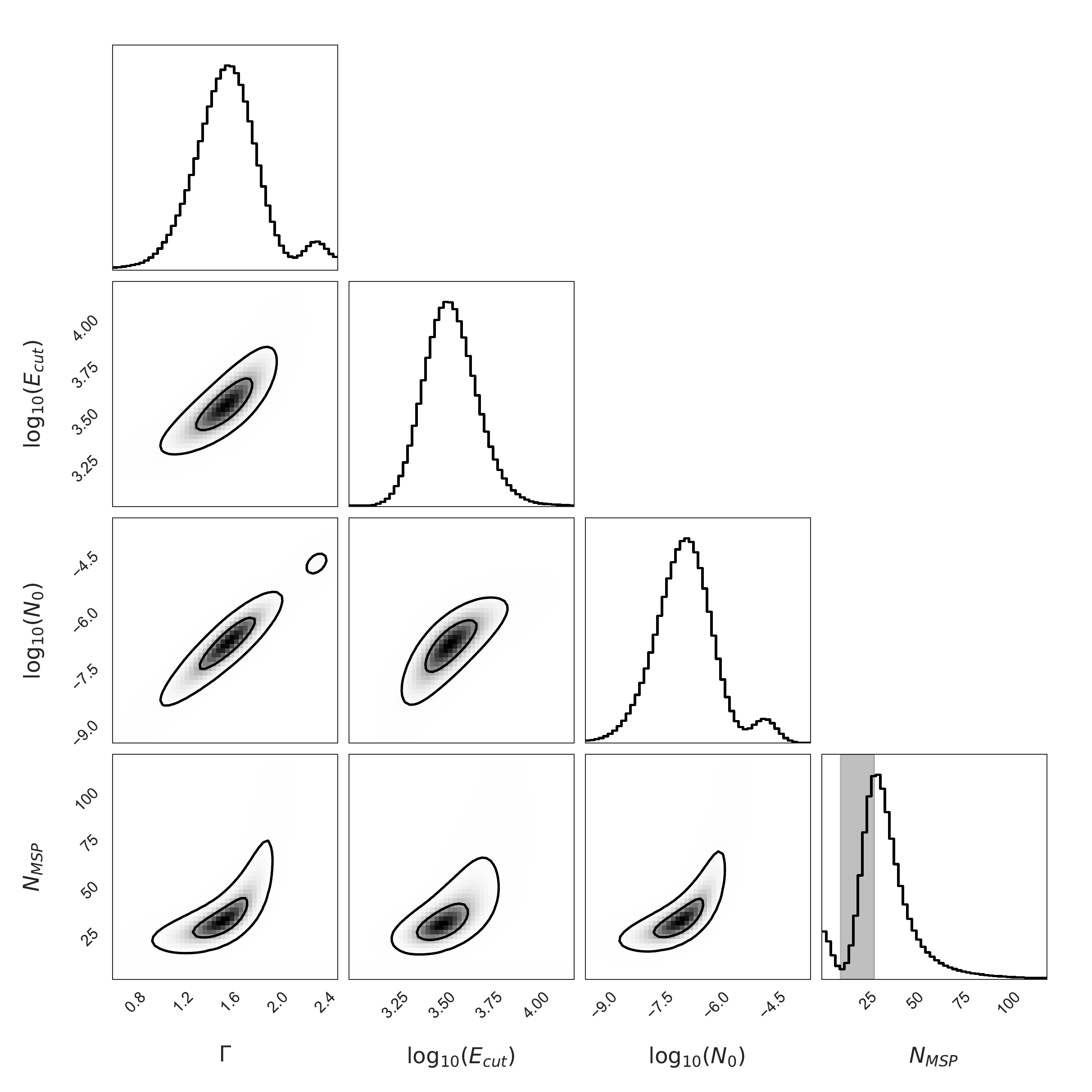}
\caption{Posterior probability distribution for the slope $\Gamma$, energy cut $E_{cut}$, and normalization $N_0$, that define the flux model for the millisecond pulsar hypothesis (Eq.~\ref{eq::exp_cutoff}). The derived number of millisecond pulsars $N_{MSP}$ has been added at the last row. The contours corresponds to the $1-$ and $2-\sigma$ confidence for our analysis, while the shaded region in the marginalized posterior for the number of MSP  corresponds to the $95\%$ confidence from the Fermi result~\cite{collaboration:2010bb}.}
\label{fig::eventscomplete}
\end{figure*}

The results for the fit are shown in Figure~\ref{fig::eventscomplete}, where we have added a last panel for the derived quantity of interest, the number of MSPs in \ocen. The flux from the MSP contribution that corresponds to the median values of the posterior is shown in Figure~\ref{fig:flux} (black dashed line).  Also shown are the fluxes from the nearby source and the background components (green and blue symbols respectively). In Figure~\ref{fig::eventscomplete} the shaded gray region in the $N_{\rm{MSP}}$ panel corresponds to the range of expected number MSPs reported by Ref.~\cite{collaboration:2010bb}. We can see those results more or less agree with our finding, although the distribution peaks at a slightly higher value, and has a long tail towards a high number of MSP.

\subsection{Dark Matter}
\label{sec:analysis_DM}
Under the dark matter annihilation scenario, the shape of the observed energy spectrum is determined by the mass of the DM particle and the model for final state of the annihilation.
There exists some theoretical motivation for considering heavy fermion pairs as the dominant annihilation final state, based upon arguments such as helicity suppression (as in the case for Majorana fermion dark matter, which includes the supersymmetric neutralino \cite{Jungman:1995df}) or considering scalar mediators, typically featuring Yukawa-like interactions with fermions (see e.g. \cite{McDonald:1993ex,Boucenna:2011hy, Profumo:2010kp,Profumo:2013yn}). However, since here we intend to fit for a $\gamma$-ray spectrum whose features indicate quite clearly a low-mass dark matter particle, it would be inappropriate to consider heavy fermions, such as top and bottom quarks, as well as pair-annihilation to gauge or Higgs bosons. Therefore we focus on the following final states: muon-antimuon ($\mu^+\mu^-$) and light quark-antiquark ($\bar{q}q$) pairs. 

To compute the number of events for these specific annihilation channels we use the following spectra.
\begin{itemize}
    \item Muons: $\frac{dN}{dy} = \frac{\alpha}{\pi} \left( \frac{1 - (1-y)^2}{y} \right) \left(\ln\left[\frac{s(1-y)}{m_\mu^2} \right]  - 1\right)$, where $y=E/m_\chi$, $\alpha$ is the fine structure constant, $s$ is the center of mass Mandelstam variable and $m_\mu$ is the muon mass. This corresponds to the use of the Altarelli-Parisi splitting function (see e.g. \cite{Bartels:2017dpb}), which is appropriate as long as the center of mass energy is much larger than the lepton mass, as is definitely the case for the range of masses of interest to us.
    \item Light quarks: $\frac{dN}{dE}= \alpha_1 \frac{E}{m_\chi}\left( \frac{E}{m_\chi} \right)^{-3/2} e^{-\alpha_2 E/m_\chi}$, which is a good approximation to simulated \gray spectra, with $\alpha_1=0.95$ and $\alpha_2=6.5$ \cite{Bergstrom:1997fj,Strigari:2006rd}. 
    
\end{itemize}

\begin{figure*}[!ht]
\centering
\subfloat[]{\includegraphics[width=0.5\linewidth]{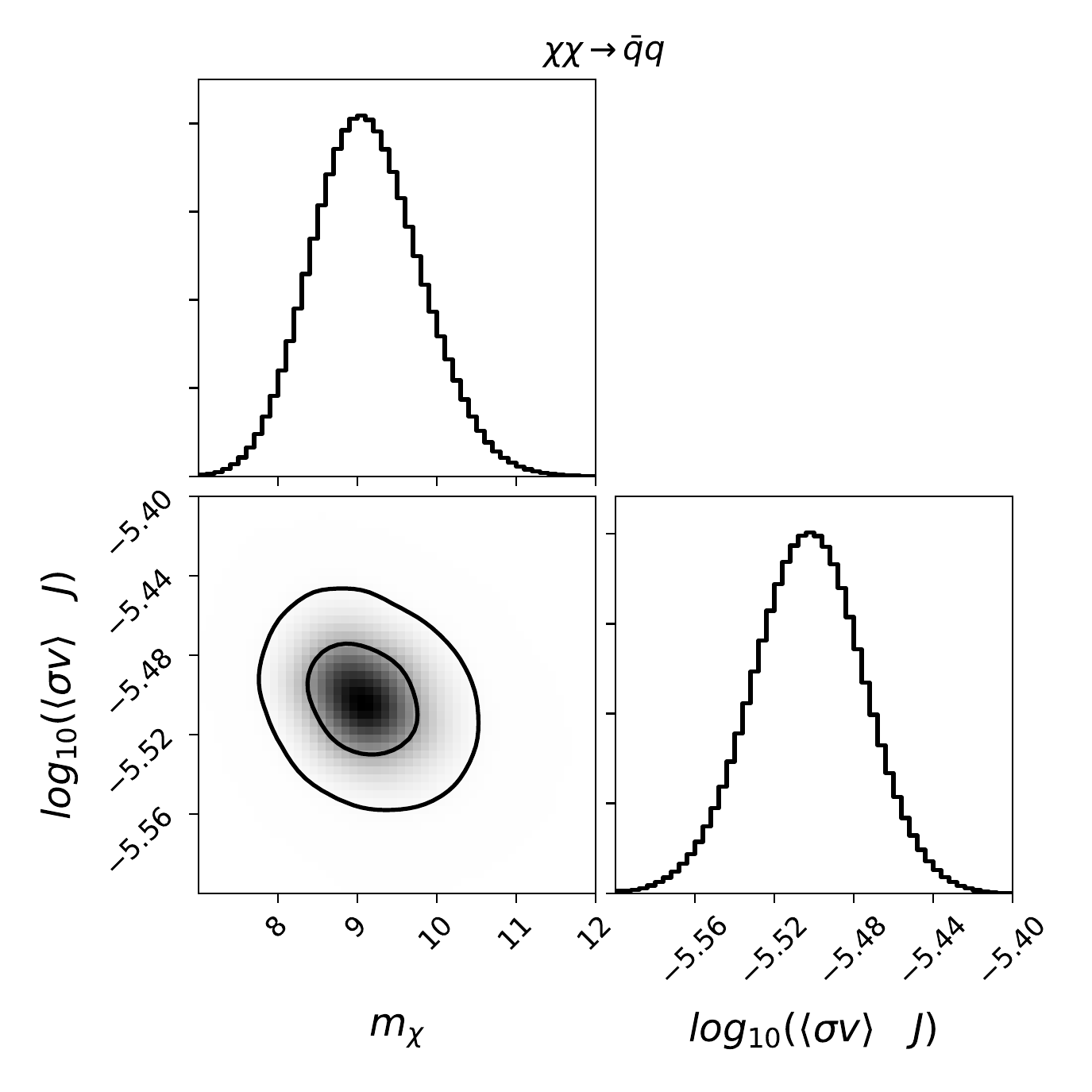}}
\subfloat[]{\includegraphics[width=0.5\linewidth]{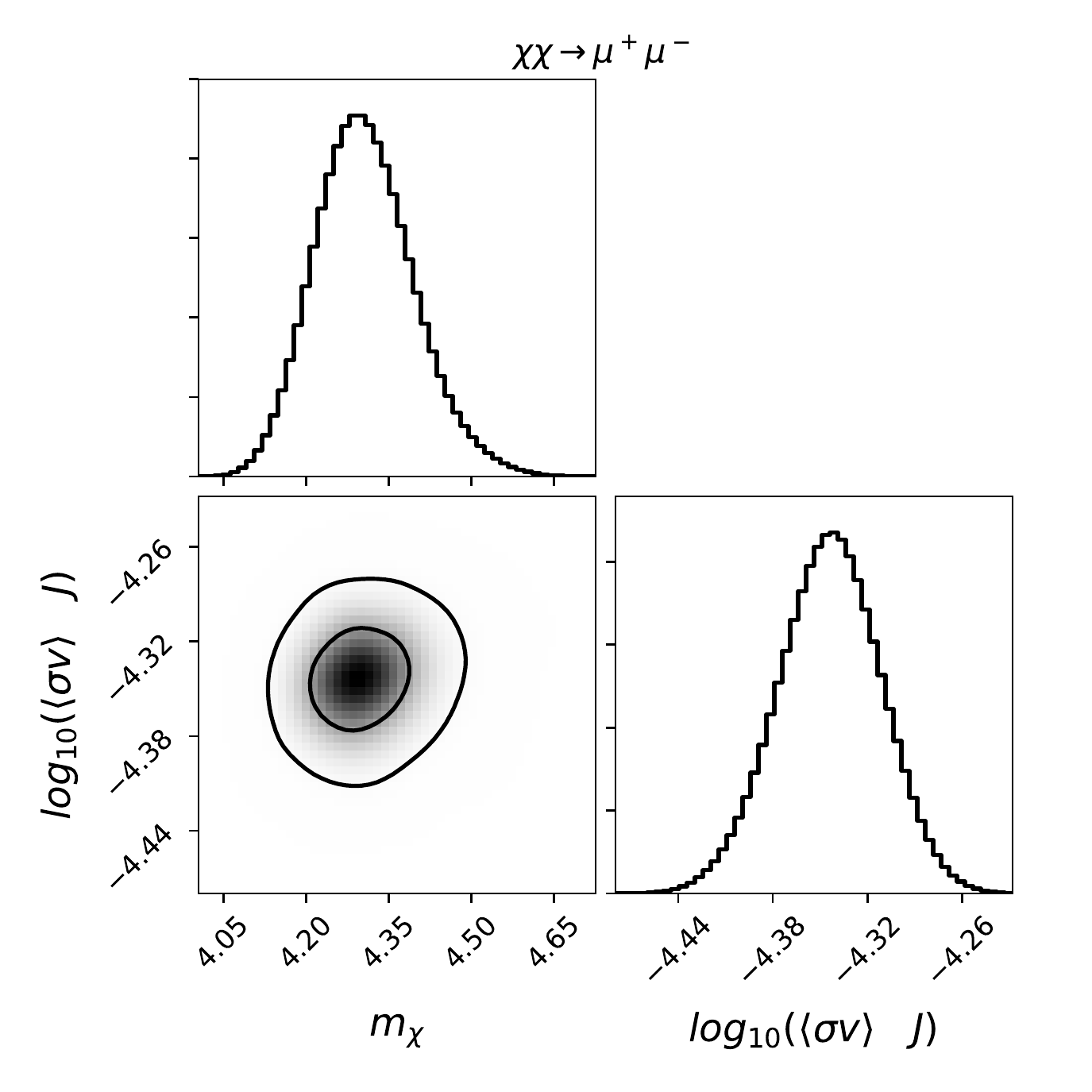}}
\caption{Posterior probability distribution for the mass of the DM particle and $\langle\sigma v\rangle J$, for the (a) $\bar{q}q$ and  (b) $\mu^+\mu^-$ annihilation channels. }
\label{fig::DM_inference}
\end{figure*}
\begin{figure*}[!ht]
\centering
\subfloat[]{\includegraphics[width=0.5\linewidth]{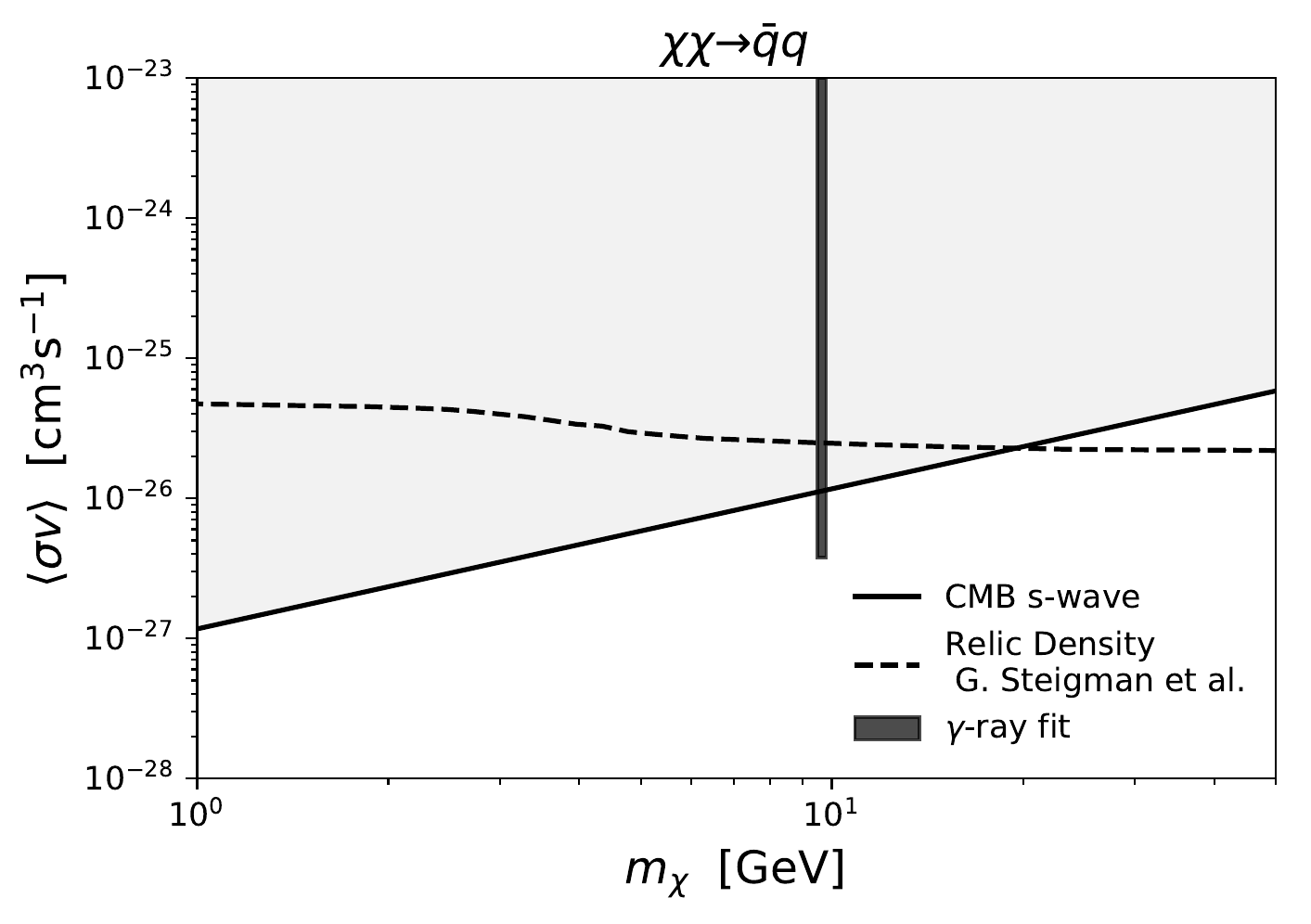}}
\subfloat[]{\includegraphics[width=0.5\linewidth]{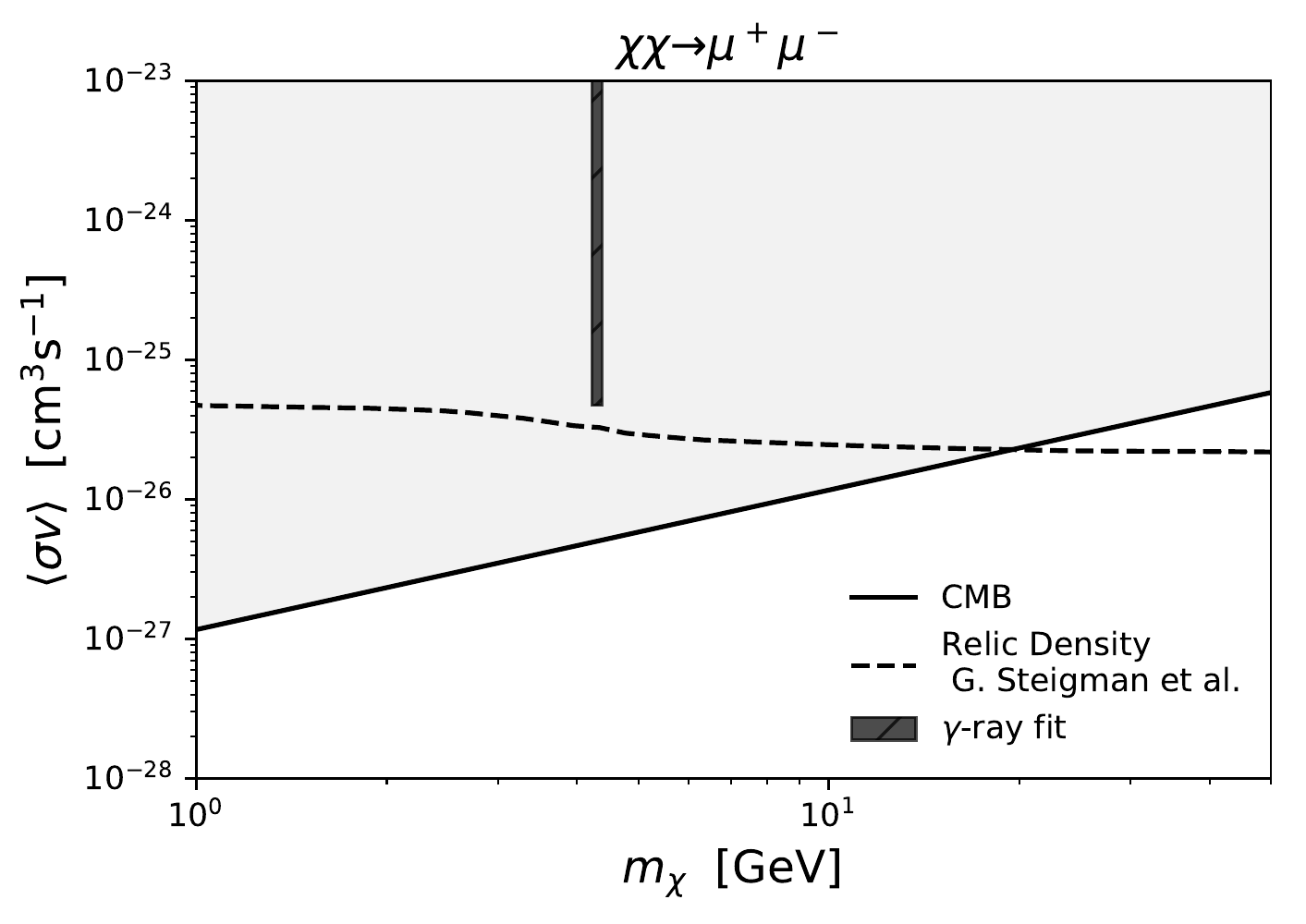}}
\caption{(a) Cross-section vs mass parameter space for DM annihilating into $\bar{q}q$ and $\mu^+\mu^-$. The dark gray vertical bar represents the 95\% confidence region that explains the observed $\gamma$-ray flux. The lower edge of the bar corresponds to our estimated maximum value of the \jfactor, see section \ref{sec::Flux_dm} for more details. The light gray shaded region is excluded by CMB observations (Eq.~\eqref{eq:cmb_constraint}). The dashed black line corresponds to the parameters that account for the total relic density.
(b) Same as (a) but for DM annihilating into $\mu^{+}\mu^{-}$. In this case there is no plausible \jfactor for \ocen that can simultaneously explain the $\gamma$-ray emission while evading the CMB constraint.}
\label{fig::cmb:contsrains}
\end{figure*}

Before detailing the MCMC analysis for these annihilation channels we summarize existing constraints. First, for s-wave annihilation, there are strong constraints on DM with particle mass from 1~MeV up to 1~GeV coming from the angular power spectrum of the Cosmic Microwave Background (CMB). An extra injection of energy from DM annihilation can ionize the neutral hydrogen during the Dark Ages, changing the effective free electron fraction left after recombination, and thus affecting the power spectrum. A thorough analysis of CMB data leads to the constraint \cite{Aghanim:2018eyx}
\begin{equation}
f_{\rm{eff}}\frac{\langle \sigma v \rangle}{m_\chi} < 3.5\times 10^{-28} \,\rm{GeV}^{-1}\rm{cm}^3 \rm{s}^{-1},
\label{eq:cmb_constraint}
\end{equation}
where $f_{\rm{eff}}$ is the effective efficiency function which quantifies the total amount of deposited energy in the hydrogen gas. While $f_{\rm{eff}}$ is redshift-dependent, it was shown that the effect on the power spectrum is the same if we consider an effective value at $z\sim 600$~\cite{Slatyer:2012yq}. We used a value of $f_{\rm{eff}}=0.3$ for both channels, as found in \cite{Slatyer:2012yq}. From Eq.~\ref{eq:cmb_constraint} we obtain an exclusion limit in the mass vs cross-section plane that is independent of the \ocen \gray analysis. This corresponds to the solid black line in Fig.~\ref{fig::cmb:contsrains}. We also show for reference the expected thermal averaged DM cross-section  as a function of the particle mass that would lead to a correct relic density for a weakly interacting massive particle (WIMP), as computed by \cite{Steigman:2012nb}.

To proceed further with the dark matter MCMC analysis we need to set priors for the DM mass and cross section. For the DM mass we choose a wide range that encloses the energies where the \gray flux is observed using the aforementioned constraints. On the other hand, whether or not we use an energy-bin dependent \jfactor or a constant one in our model (Eq.~\ref{eq::convolution}), it is clear from Eq.~\ref{eq::flux} that there is a degeneracy between the \jfactor and the cross section. When using a constant value for $J$, we can combine the two into a single parameter $\left \langle \sigma v \right \rangle J$.  The prior on this parameter combination is not so evident. By choosing a $\log_{10}(J/\mathrm{GeV^2 cm^{-5}})=21$, which we find to be a plausible maximum value for \ocen based on Figure \ref{fig:j_factor}, we consider a prior for the cross section in the range $10^{-31}<\left \langle \sigma v \right \rangle/\mathrm{cm^3 s^{-1}} < 10^{-21}$. This would result in a range of $-10<\log_{10}(\left \langle \sigma v \right \rangle J)<0$. For smaller values of the \jfactor, the limits of the $\left \langle \sigma v \right \rangle J$ product would be highly constrained by the CMB. With this estimate we chose a large enough prior to include a wide  parameter space in the $\left \langle \sigma v \right \rangle J$ Vs mass plane:

\begin{itemize}
    \item $0.5<m_\chi /{\text{GeV}}<20$,
    \item \! $-20 < \log_{10}(\left \langle \sigma v \right \rangle J /\text{GeV}^2\text{cm}^{-2}\text{s}^{-1})<0$,
\end{itemize}

For the nuisance parameters of the nearby source we used the same Gaussian priors as in the MSP case. In Figure~\ref{fig::DM_inference} we show the results of the MCMC for the $\mu^+\mu_-$ and $\bar{q}q$ annihilation channels.

\begin{table}[!t]
\renewcommand{\arraystretch}{1.5}
\begin{center}
\begin{tabular}{ c c c } 
 & \boldmath{$\mu^+\mu^-$} & \boldmath{$\bar{q}q$} \\
\hline
\boldmath{$m_\chi$} & $4.30^{+0.09}_{-0.08}$ & $9.10^{+0.69}_{-0.62}$\\
\boldmath{$\log_{10}(\left \langle \sigma v \right \rangle J)$} & $-4.34^{+0.03}_{-0.03}$ & $-5.50^{+0.03}_{-0.03}$\\
AIC & $179.4$ & $47.4$ \\
\hline
\end{tabular}
\caption{Median values found by the parameter inference for two different dark matter annihilation channels (see figure~\ref{fig::DM_inference}) and the corresponding Akaike information criterion parameter (AIC). The error bars are at a $95\%$ confidence.  }
\end{center}
\label{table::DM_parameters}
\end{table}

In Figure~\ref{fig::cmb:contsrains} we present the CMB constraints (black solid line), the relic density (black dashed line),  and the results of the MCMC analysis (gray vertical shaded region), where the lower limit on \sigmav corresponds the reference value of $\log_{10}J=21$. As we can see, for our choice of the \jfactor, annihilation into muons is excluded by Planck constraints, in the s-wave approximation,  whilst the case of $\bar{q}q$ shows a small allowed window. A better determination of the \jfactor will help reduce further the allowed parameter space. On the other hand, this could also be interpreted as an upper limit for the \jfactor for a given $\langle \sigma v \rangle$.

 As we claimed in Section~\ref{sec::Flux_dm}, the formal way to proceed in the analysis would be to include the convolution of the PSF with the \jfactor. This can be done by introducing such computation into the MCMC, leading to an exploration of a six-dimensional parameter space, the thermal averaged cross section $\langle \sigma v \rangle$, the DM density profile slope $\gamma$, the scale radius $r_s$,  the particle mass $m_{\chi}$, and the two nuisance parameters associated to the nearby source. The computation of the aforementioned convolution is computationally demanding, but the actual limitation is that there is a clear degeneracy between the first three parameters. Therefore, in this work we argue that using a constant \jfactor multiplied by the PSF, instead of the convolution will not alter our conclusions. As seen on Figure \ref{fig:j_factor}, the convolution for the higher energy bins approaches the value the non-convolved \jfactor. In order to quantify the error introduced by taking a constant \jfactor we took a value of $r_s=7$ pc and $\gamma=0.5$, computed the convolution for all energy bins, and performed the MCMC analysis over the rest of the parameters (\sigmav, $m_\chi$, $N_s$ and $\alpha_s$). We then compared the result with an identical procedure but fixing the \jfactor to the non-convolved value for this combination of $r_s$ and $\gamma$, $log_{10}(J)=19.3$.  We found that for $\bar{q}q$ the value of the inferred  DM particle mass changes approximately $\sim 17\%$, while in  the case of DM annihilating into $\mu^+ \mu^-$ leaves the posterior unchanged.  This can be attributed to the fact that the muon spectrum does not contribute strongly to the first energy bins, which have the largest discrepancy once the convolution is taken into account. In both channels the posteriors on \sigmav are unaffected.
 
\section{Discussion}
\label{sec:Discussion}

The motivation for this work is connected to the following observations: 
 
(i) Given that \ocen's formation is not entirely understood and that its present-day properties (such as metallicity) differentiate it from other GCs, it is possible that it is the remnant of a dwarf galaxy whose DM halo has been tidally disrupted~\cite{1996ApJ...462..241N,Hilker:2000hr,Gnedin:2002un,Carraro:2000ji,Marcolini:2007kz,Romano:2007kh,2013ApJ...762...36J,Lee:2009gg,Bekki:2005rx,Rey:2003wk,Fellhauer:2004gc,Majewski:1999fw,1988IAUS..126..603Z,2004MNRAS.350.1141T}. 

(ii) Recent studies on \ocen's internal kinematics suggest a larger mass-to-light ratio than expected from a stellar population analysis.

(iii) There is a well-identified $\gamma$-ray emission associated to it that so far cannot be definitively identified with known astrophysical objects, MSPs in particular.

Under these premises, we studied \ocen by fitting the $\gamma$-ray observed flux using two different models for its origin: unresolved MSPs and DM annihilation.

\subsection{MSP}
\label{sec:Discussion:MSP}
In the MSP scenario the quantity of interest is the number of MSPs  inferred from the observed flux. Taking a simple approach, we concluded that a median value of $\sim 30$ MSPs are needed to explain the observed $\gamma$-ray emission, slightly higher than the expected number reported in \cite{collaboration:2010bb}. This can be understood as current data indicates a higher energy cut off  $E_{\rm{cut}}$ and harder spectral index $\Gamma$ than reported in Ref.~\cite{collaboration:2010bb}. Nonetheless, the number of MSPs obtained here is consistent with the previous work of \cite{deMenezes:2018ilq} who estimated the number using a similar approach to \cite{collaboration:2010bb}. However, it has been discussed whether the estimate of the MSP population computed through Eq.~\eqref{eq::no_events_pulsar} is reliable \cite{Cholis:2014noa}; it could be that the luminosity of \ocen comes from a single very luminous pulsar or it could be the contribution of many fainter ones. The fact that there are $\sim 30$ MSP candidates \cite{Henleywillis:2018ufu} in X-ray points to the plausible explanation of multiple MSP that also produce the $\gamma$-ray emission. With stringent constraints on the number of MSPs a deeper analysis on the luminosity function will be needed. 

\subsection{Dark Matter}
\label{sec:Discussion:DM}
Another possible scenario is that \ocen is the remnant of a dwarf galaxy, leading to the possibility that dark matter makes a (subdominant) contribution to \ocen's mass. In this work we make a first attempt to use the observed $\gamma$-ray flux associated with \ocen to constrain the DM particle properties, and overall DM content itself. Given the energy range of the observed flux, we explored the possibility of it being produced by dark matter particles annihilating into $\mu^+ \mu^-$ and $\bar{q}q$. For the analysis we treated the product of the astrophysical \jfactor and the thermal averaged cross section as a free parameter $J \left \langle \sigma v \right \rangle$ along with the DM particle mass $m_\chi$. The constraints on such parameters, found on the MCMC analysis, are reported in Table \ref{table::DM_parameters}. Figure~\ref{fig::cmb:contsrains} summarizes our results. 

 With the posteriors in hand, we compare the analysis results with current DM annihilation constraints. In Figure \ref{fig::cmb:contsrains}, the light shaded region corresponds to the thermal-averaged cross section values that are excluded by the latest Planck CMB constraints (for s-wave annihilation). The dark gray vertical band corresponds to the allowed region at the 95\% confidence level, for the DM mass, found in this work. The lower limit on the cross section is set by the largest \jfactor we consider possible for \ocen. For the case of DM annihilating into $\bar{q}q$, there is an allowed range of \jfactor values that will be consistent with both constraints on the maximum amount of DM in \ocen, and the $\gamma$-ray flux. Conversely, the case of DM annihilating into $\mu^+ \mu^-$, constraints are more stringent. All of the parameter space is already excluded by CMB, meaning that it will require a very small cross-section and/or a very large \jfactor (which translates into very compact objects) to explain the $\gamma$-ray flux. Note that taking a p-wave annihilation instead on s-wave relaxes CMB constrains due to the fact that the p-wave DM annihilation does not contribute importantly to the thermal history of the Universe until low redshift.

Despite the fact that only for the $\bar{q}q$ case there is a small window of the allowed parameter space, an assumption that the spectra is produced by multiple contributors can in principle change the values of the mass and cross-section needed to explain the data, possibly lowering the cross section needed to reproduce the photon flux. Recent work assessed the possibility that the \gray emission from another globular cluster, 47 Tuc, could be explained by both MSP and DM annihilation. According to \cite{Brown:2018pwq} the observed photon flux is better fit by a model that includes DM than by a model of MSPs alone, excluding the latter with more than five sigma significance. While \cite{Bartels:2018qgr} argue there is no significant evidence for a DM component if the variance in the spectrum of a population of MSPs is included in the analysis. This stresses the importance of taking into account the uncertainties in the MSP contribution to the photon flux, specially for mixture models (MSPs and DM).  In this work, however, we focused on separate limits for each DM annihilation channel in an effort to gain insight not only on possible explanations to \ocen's $\gamma$-ray emission, but also to obtain conservative limits on the DM content of \ocen under the tidally disrupted dwarf hypothesis.

The results presented were analyzed by taking a $0.2\deg$ ROI. Nonetheless, for both the MSP and DM hypotheses we repeated our analysis using a $0.15\deg$ and a $0.5\deg$ ROI to check for consistency. The posteriors for all three angular regions overlap within $1 \sigma$ with the three different data sets.

The predicted $\gamma$-ray flux for each DM model and for the MSP model is shown in Figure~\ref{fig:flux}, based on the values presented in Tables~\ref{table::MSP_parameters} and~\ref{table::DM_parameters}. We see that, qualitatively, the three models are a relatively good fit to the data. A quantitative comparison using the AIC factor, shown in the aforementioned tables, leads to the conclusion that the  MSP explanation provides a better fit than the DM models. As in \cite{Brown:2018pwq, Bartels:2018qgr} a joint analysis with both MSPs and DM could  be performed. After an eventual confirmation of a population of MSPs in \ocen, they could be included as an additional background in the DM annihilation scenario. In this case, the DM constraints would become much more stringent. Because of \ocen's potentially large \jfactor, the DM constraints may eventually be competitive with those from dwarf spheroidal galaxies. To reach this level we also need a better knowledge of the dark matter density distribution within \ocen. This can be explored in future dedicated analysis of \ocen's stellar kinematics.

To summarize, here we have addressed for the first time the possibility that the $\gamma$-ray emission from \ocen can be explained by DM annihilation. Such models are quite constrained. Constraints can also be given in terms of the maximum content of DM in \ocen. Further studies of \ocen will shed light on the origin of this peculiar system. Dedicated analysis of the \jfactor will help narrow down the possibilities for DM to explain the observed $\gamma$-ray flux and the confirmation of the existence of MSPs (or other astrophysical sources) will allow stronger constraints on the DM content and properties of \ocen.

\section*{Acknowledgements}
This work was partially funded by a UCMEXUS-CONACYT collaborative project and CONACYT project 28689. AXGM acknowledges support from Cátedras CONACYT.  SP is partly supported by the U.S. Department of Energy grant number de-sc0010107.OV acknowledges support from the PAPIIT-UNAM grant IN112518.

\appendix
\section{Integrated PSF for nearby point sources}
\label{app:PSF}
As stated in Section \ref{sec:Flux}, to account for the number of events we must consider the instrument response and in particular the information contained in the point spread function (PSF). The energy-dependent probability of detecting a photon within the ROI is computed by integrating over the ROI,
\begin{equation}
{\rm PSF}(E)=\int_{\Delta \Omega} {\rm PSF}(\theta|E)\hspace{1mm} d \Omega,
\end{equation}
where $\theta$ is the angle between the point within the ROI and the \gray source.

For the small angles under consideration we can take the sky as a flat plane and integrate the PSF in rectangular coordinates. The standard integration for \ocen is straightforward and has no major complication. However, when we perform the integration for the nearby source, we must integrate its PSF also in a circle centered on \ocen, this can be achieved by
\begin{equation}
\label{eq::int_PSF}
{\rm PSF}(E)=\int_{-\theta}^{\theta}dx \int_{-\sqrt{\theta^2 - x^2}}^{\sqrt{\theta^2 - x^2}}dy {\rm PSF}\left(\sqrt{(D-x)^2 + y^2}|E\right),
\end{equation}
where $\theta$ is radius of the ROI and $D=0.63^\circ$ is the angular separation between the source and \ocen.

\section{Convolution of the Point Spread Function and the \jfactor}
\label{sec::convolution}

To compute the differential flux we must convolve the $J$-profile and the PSF as in Eq. \eqref{eq::convolution}. We make use of the very efficient procedure introduced in \cite{Geringer-Sameth:2014qqa}. Just as in the last section, we take the source to be in a 2D rectangular plane where the angular separation is approximated by the Euclidean distance. The convolution then has to be done through a 2D integral in the $xy$ plane. We compute the convolution very efficiently by going to Fourier space and taking advantage of the circular symmetry of both the $J$-profile and PSF. Let $(J*{\rm PSF})(x,y)$ be denoted by $f(x,y)$. Then
\begin{equation}
f(x,y)=\int dx' dy' \, J(x',y') \, {\rm PSF}(x-x',y-y').   
\end{equation}
The Fourier transform of $f$ is a function only of the magnitude of the $k$-vector $\hat{f}(k_x,k_y)=\hat{f}\left(k= \sqrt{k_x^2+k_y^2}\right)$ and takes the form
\begin{equation}
\label{eq::Hankel_transform}
\begin{aligned}
\hat{f}(k)= \left(\int_0^\infty 2 \pi r \frac{dJ}{d\Omega}(r)\jmath_0 (kr) dr\right)\\\times \left( \int_0^\infty 2 \pi r {\rm PSF}(r,E) \jmath_0 (kr) dr\right), 
\end{aligned}
\end{equation}
where $\jmath_0$ is the Bessel function of zero order and $dJ/d\Omega$ is defined in Eq.~\ref{eq::Jfactor} ($r$ being the Euclidean approximation to angle $\theta$). The two terms in~\eqref{eq::Hankel_transform} (except for a factor of 2$\pi$) are the zeroth-order Hankel transforms, which can be performed efficiently using the FFTlog algorithm \cite{Hamilton:1999uv} and the python package \texttt{mcfit}. We followed the description of the algorithm described in Ref.~\cite{Geringer-Sameth:2014qqa} to avoid numerical issues when using mcfit and we used the default parameters given by the package functions.
\bibliographystyle{unsrt}
\bibliography{bib.bib}

\end{document}